%% file: main.tex
\documentclass[journal]{IEEEtran}

\usepackage{graphicx}
\usepackage{bm}
\usepackage{enumerate}
\usepackage{float}
\usepackage{multicol}
\usepackage{color}
\usepackage{url}
\usepackage{cite}
\usepackage{amsmath}
\usepackage{amsthm}
\usepackage{amssymb}
\usepackage{array}
\usepackage{multirow}
\usepackage[table,xcdraw]{xcolor}

\usepackage{flushend}

\renewenvironment{IEEEbiography}[1]
  {\IEEEbiographynophoto{#1}}
  {\endIEEEbiographynophoto}

\ifCLASSOPTIONcompsoc
  \usepackage[caption=false,font=normalsize,labelfont=sf,textfont=sf]{subfig}
\else
  \usepackage[caption=false,font=footnotesize]{subfig}
\fi

\usepackage{acronym}

\PassOptionsToPackage{hyphens}{url}
\usepackage{hyperref}

\input{main_acro.tex}

\begin{document}

\begin{titlepage}
\vspace{5cm}
\Huge\noindent\textbf{IEEE Copyright Notice}\\
\vspace{3cm}

\large\noindent\copyright 2022 IEEE. Personal use of this material is permitted. Permission from IEEE must be obtained for all other uses, in any current or future media, including reprinting/republishing this material for advertising or promotional purposes, creating new collective works, for resale or redistribution to servers or lists, or reuse of any copyrighted component of this work in other works.

\end{titlepage}

\captionsetup{belowskip=0pt,aboveskip=0pt}

\title{High-Rate Uninterrupted Internet-of-Vehicle Communications in Highways: Dynamic Blockage Avoidance and CSIT Acquisition}

\author{Hao~Guo,~\IEEEmembership{Graduate~Student~Member,~IEEE},
        Behrooz~Makki,~\IEEEmembership{Senior~Member,~IEEE},\\
        Mohamed-Slim Alouini,~\IEEEmembership{Fellow,~IEEE},
        and Tommy~Svensson,~\IEEEmembership{Senior~Member,~IEEE}}

\maketitle

\begin{abstract}
In future wireless networks, one of the use-cases of interest is Internet-of-vehicles (IoV). Here, IoV refers to two different functionalities, namely, serving the in-vehicle users and supporting the connected-vehicle functionalities, where both can be well provided by the transceivers installed on top of vehicles. Such dual functionality of on-vehicle transceivers implies strict rate and reliability
requirements, for which one may need to communicate at millimeter wave (mmW) frequencies. However, IoV communication at mmW requires up-to-date channel state information (CSI) and blockage avoidance. In this article, we incorporate the recently proposed concept of predictor antennas (PAs) into a \textit{large-scale cooperative PA (LSCPA)} setup where both  temporal blockages and  CSI out-dating are avoided via base stations (BSs)/vehicles cooperation. Summarizing the ongoing standardization progress enabling IoV communications, we present the potentials and challenges of the LSCPA setup, and compare the effect of cooperative and non-cooperative schemes on the performance of IoV links. As we show, BSs cooperation and blockage/CSI prediction can boost the performance of IoV links remarkably.
\end{abstract}

\IEEEpeerreviewmaketitle


\section{Introduction}
Suppose that, sitting in an autonomous drive vehicle driving in a highway, you are watching Game of Thrones, and your smart phone gets disconnected. In ten years, this will bring huge dissatisfaction for the users, leading to poor rating for the network provider. Particularly, in the 6G era, users in their own cars expect the same \ac{QoS} as they have at home. Thus, cellular networks will face with high data rate/reliability demands driven by in-vehicle users running, e.g., 4K/8K video streaming applications. This is a part of the \ac{IoV} concept, for which high-rate uninterrupted connections are required. 

Such reliable \textit{mobile} \ac{IoV} links are used not only to serve  in-vehicle users but also for  connected-vehicle/autonomous drive use-cases relying on \ac{V2I}, in general V2X (X: everything), communications. Currently, connected automated drive is a large business market with more than \$130 billion revenue in 2019, and is expected to garner more than \$500 billion by 2026. Thanks to V2X communications, traffic information becomes more precise, improving the traffic efficiency/safety and reducing CO2 emissions. However, V2X links have strict requirements. For example, 99.99\% reliability is required for network-assisted pedestrian protection, and considerably higher reliability is foreseen with Levels 4/5 of self-driving vehicles being commercially available  by 2030 \cite{mikael2021}.

Along with connected-vehicle related communications, the antennas mounted on top of the vehicles can be utilized as intermediate \acp{MR} connecting the \acp{BS} and the in-vehicle users. This is motivated by the fact that 1) compared to typical \acp{UE}, such \acp{OVT} may be equipped with more antennas/capable of advanced signal processing techniques. Also, 2) compared to direct \ac{BS} to in-vehicle links, the \ac{MR} implementation eliminates the vehicle penetration loss, which is measured to be around 25 dB at 2.4 GHz \cite{guo2021predictor} and even higher losses are foreseeable at higher frequencies. Such dual functionality of \ac{IoV} transceivers, however, increases the rate/reliability requirements of the \ac{BS}-\ac{OVT} links even further. 

To increase the data rate, different methods are considered, among which network densification and \ac{mmW} communications are the dominant ones. Network densification refers to the deployment of multiple \acp{BS} of different types in an area. Particularly, it is expected that soon small cells will be densely deployed to assist the existing macro \acp{BS}. On the other hand, \ac{mmW} transmission offers wide bands and simplifies multi-antenna communications. However, \ac{mmW} transmission experiences poor propagation characteristics and has high free-space path loss. For these reasons, it is expected that \ac{mmW}-enabled small cells will be deployed at low heights, e.g., on lamp post, with a short coverage range. With a low antenna height, however, the probability of blockage increases, which reduces the achievable rate of the \ac{mmW} links significantly. Therefore, to guarantee uninterrupted high-rate communications, it is preferred to avoid blockage.

With stationary networks, a large part of static blockages are avoided via deployment
optimization. Also, one can avoid (semi-)static blockages  via resource association \cite{liu2020user}, \ac{CP} transmission \cite{fang2020hybrid} or the incorporation of relays \cite{madapatha2020integrated}/intelligent reflecting surfaces \cite{zhou2021sto}. Alternatively, back-up \ac{NLoS} links can be found  during the initial beam training phase and, if a \ac{LoS} link is blocked, the connection can switch to the back-up link(s) \cite{tunc2020}.

The problem, however, becomes more challenging in moving, e.g., \ac{V2I}, networks with high-speed vehicles driving in highways. First, deployment optimization does not help in moving networks. Second, the back-up \ac{NLoS} solution is not of interest because, 1) good reflectors (building, etc.) are rare in highways, and 2) compared to urban scenario with rich scattering, such back-up links sustain for a shorter period \cite{tunc2020}. Third, considering low-height small cells in highways, the blockers are mainly buses and trucks passing by the cars. Thus, probabilistic blockage prediction or (machine learning-based) deployment learning methods may not cope well with the network dynamics at high speeds/frequencies. This is specially because not only the blockage results in excessive \ac{SINR} degradation, but also the system performance is considerably affected by channel aging phenomena where with high speeds the \ac{CSIT} soon becomes inaccurate. 

In this article, we investigate different methods of dynamic blockage avoidance and \ac{CSIT} prediction in \ac{IoV} networks. We concentrate on a highway scenario where \acp{OVT}, either used for connected-vehicle applications or as an MR, receive high-rate uninterrupted streams transmitted by
\ac{CP} \acp{BS}. We incorporate the recently proposed \ac{PA} concept \cite{guo2021predictor,phan2018WSAadaptive} into a \ac{LSCPA} setup with \ac{CP} communications among \acp{BS}, and utilize the information provided by
different vehicles to avoid not only temporal blockages but also the \ac{CSIT} out-dating. We compare the performance of various alternative technologies at different frequencies and levels of adaptivity/cooperation, and show that there is not a single method providing the best reliability/throughput.
Instead, a combination of different methods gives the best performance. Finally, we extend the PA concept
to the cases with \ac{MIMO} communications, and study its performance with
different levels of CSI. As we show, joint dynamic blockage and \ac{CSIT} prediction improves the throughput and reliability of \ac{IoV} links by orders of magnitude.


\section{MmW IoV Communications in Highways}
Initial intelligent transportation systems were based on \ac{DSRC} standards, e.g., IEEE 802.11p/\ac{DSRC}, supporting data rate in the order of 10 Mbps. On the other hand, connected vehicles are currently equipped with hundreds of sensors where, for instance, \ac{LiDAR} system alone may require data rates up to 100 Mbps  for blind spot removal \cite{tassi2017modeling}. Obviously, \ac{DSRC}-based systems are not enough for such applications. Also, although \ac{LTE-A}-based solutions boost the  rates to up to 100 Mbps  \cite{tassi2016analysis}, we still need higher rates. This is particularly because, with the \ac{IoV} concept, the \acp{OVT} not only provide connected-vehicle communications but also work as \acp{MR}. 

With this background and motivated by the 5G progress in \ac{mmW} communications, multi-antenna \ac{mmW} transmission is a powerful candidate for \ac{IoV} providing massive bandwidth and high-gain \ac{BF}. MmW systems are typically considered in static networks, e.g., wireless backhaul \cite{madapatha2020integrated}. With mobility, however, the availability of the \ac{mmW}-based narrow \ac{BF} systems is prone to different challenges including \ac{CSIT} inaccuracy, beam misalignment and blockage. Particularly, with a blockage the path loss exponent increases from $\sim 2.8$ in the cases with \ac{LoS} connections to $\sim3.9$ in \ac{NLoS} communication, and outdated \ac{CSIT}/out-of-beam signal reception lead to significant \ac{SINR} drop.  

\begin{figure*}
\centering
  \includegraphics[width=1.8\columnwidth]{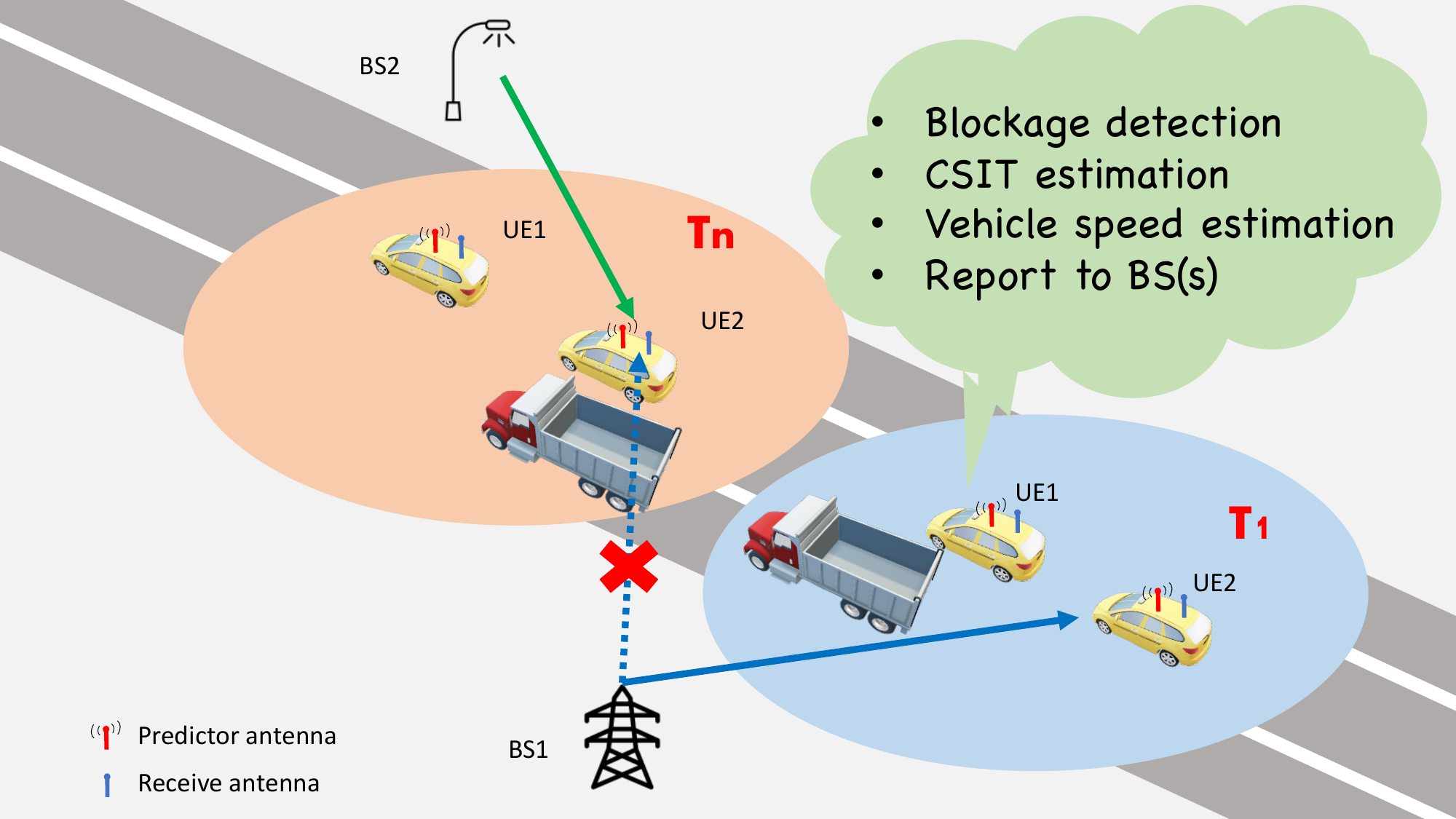}\\
\caption{The proposed \ac{LSCPA} system. The information provided by the front vehicle ($\text{UE}_1$) at time $T_1$ is used to predict both blockage and small-scale fading for data transmission to the following vehicles ($\text{UE}_2$) at time $T_n$ via a, possibly different, BS. }
\label{fig1}
\end{figure*}

The key features of highway environments, however, help to solve the dynamic blockage/channel aging issues to some extent; In a highway,  connected vehicles are likely to drive along the same set of lanes with controlled speeds. Here, an example is platooning setup, i.e., a group of vehicles traveling closely together with speed/direction controlled by the lead vehicle. Moreover, as illustrated in Fig. \ref{fig1}, in a highway, slow large vehicles (e.g., buses, trucks) travel typically in the outermost lanes, while high-speed vehicles drive along the innermost lanes. Then, temporal blockage occurs if a large vehicle drives between a user and its serving \ac{BS} located on lamp posts along the highway. In this case, as we explain in the following, one can use channel prediction and  \ac{CP} communication/information exchange between the vehicles and \acp{BS} not only to avoid channel aging  but also the temporal blockages. 

Few works study \ac{mmW} communication in highways. The interesting work of  \cite{tassi2017modeling} develops theoretical models to describe the \ac{DL} \ac{V2I} network in the presence of blockage and  perfect \ac{CSIT}. Also, \cite{yi2019modeling} designs a model for \ac{mmW} V2X networks with platooning. Finally, \cite{eshteiwi2020impact} develops a V2V \ac{CP} network assuming perfect \ac{CSIT}. 

\section{Large-scale Cooperative Predictor Antenna}
One option for \ac{CSIT} prediction at high speeds is to use Kalman-filter based channel estimation (see \cite{guo2021predictor} for a review of different small-scale \ac{CSIT} prediction schemes at high speeds). Using the \ac{PA} concept for small-scale \ac{CSIT} acquisition in \ac{MR} systems has been studied in, e.g.,  \cite{guo2021predictor,phan2018WSAadaptive}. Here,  the idea is to deploy a \ac{PA} at the front of a vehicle which is used to estimate  the channel quality that is observed by \acp{RA}, aligned behind the \ac{PA}, when they reach the same point as the \ac{PA} (see Fig. \ref{fig1}). Specially, in the first time slot, the \ac{PA} estimates the channel and sends CSI back to the \ac{BS}. In the second time slot, while the \ac{BS} needs time to transmit to the \ac{RA}, the vehicle moves forward. The \ac{BS} will
have good estimate of the \ac{BS}-\ac{RA} link CSI, if the \ac{RA} reaches the same point as the \ac{PA}.   As shown in \cite{guo2021predictor,phan2018WSAadaptive}, the \ac{PA} setup can boost the prediction horizon of the Kalman prediction-based schemes with 0.1-0.3 wavelengths prediction horizon to up to 3 wavelengths at 2.68 GHz and a velocity of 45-50 km/h.

The feasibility of the \ac{PA} concept in a single vehicle has been verified by testbed experiments. For example, in 2014, Dresden, Germany, we showed that the cross-correlation between the \ac{PA} and the \ac{RA} maintains  high  (more than 97\%) for up to three times the wavelength \cite{Jamaly2014EuCAPanalysis}. Also,  in 2018,  \cite{phan2018WSAadaptive} performed channel measurements in Stuttgart, Germany, where a 2.18 GHz system was set up with 64 antennas at the \ac{BS} and 2 monopole antennas at the vehicle. Assisted by the \ac{PA}, such a \ac{MIMO} \ac{DL} system could reach the ideal \ac{BF} gain even in \ac{NLoS} scenarios. Interestingly,  \cite{phan2018WSAadaptive}  showed that, assisted by the \ac{PA}, applying zero-forcing transmission at the \ac{BS} to two spatially multiplexed cars leads to 20 to 30 dB \ac{SINR} gain.

The current \ac{PA} setups are designed for 1) communication between a single vehicle and a \ac{BS}, 2) \ac{NCP} communications, and 3) predicting the small-scale fading. Here, we develop the \ac{LSCPA} concept  for both blockage and channel aging avoidance  in a \ac{CP} fashion as described in the following (see Fig. \ref{fig1}).

With \ac{LSCPA}, each vehicle is equipped with, possibly, one set of \ac{PA}/\ac{RA} antennas on top of the vehicle. Here, as illustrated in Fig. \ref{fig1}, while the \ac{PA} of each vehicle can work as a small-scale predictor for its own RA, as in the existing \ac{PA} setups \cite{guo2021predictor,phan2018WSAadaptive}, the antennas on top of the front vehicle can enable reliable high-rate data transmission to the vehicles behind from different perspectives: 
\begin{itemize}
\item  If the front vehicle, i.e., $\text{UE}_1$ in Fig. \ref{fig1}, detects a blockage, e.g., by a truck passing by, it estimates the speed of the blocker and $\text{UE}_2$. Then, $\text{UE}_1$  informs one (or, multiple) \ac{BS}(s) about both the instantaneous \ac{CSIT} of the location and the speed of the blocker and $\text{UE}_2$. Knowing the speeds, the \ac{BS}(s) predict the slots, e.g., Slot $T_n$ in Fig. \ref{fig1}, when $\text{UE}_2$ will be blocked by the truck and, in those slots, we may switch to a different \ac{BS} to serve $\text{UE}_2$ blockage-free.
\item  Without blockage, the instantaneous \ac{CSIT} provided by the \ac{PA} of the front vehicle can assist the data transmission to the second vehicle. Here, utilizing the \ac{CSIT} of $\text{UE}_1$, both antennas of $\text{UE}_2$ are used for data reception when $\text{UE}_2$ reaches the same position as where the antenna of $\text{UE}_1$ was sending pilots, and no extra pilot transmission is required for $\text{UE}_2$. Alternatively, the \ac{PA} of the second vehicle can also be used for \ac{CSIT} acquisition and such an information is combined with that provided by the first vehicle to improve the \ac{CSIT} accuracy. Here, one can consider different methods to combine the \ac{CSIT} provided by different antennas. 
\end{itemize}

Thus, the cooperation is on sharing the speed/location of the vehicles, the \ac{CSIT} as well as handover between  \acp{BS}. In this way,  not only the small-scale fading prediction can be improved, but also  the temporal blockages are detected and avoided. The former can be utilized to compensate for the channel aging effect, while the latter enables dynamic blockage avoidance by enabling a vehicle to switch to another \ac{BS} when predicted that a possible blockage may occur to the link from the vehicle to its currently allocated \ac{BS}. Also,  our setup  simplifies  \ac{HARQ}-based retransmissions and reduces the number of retransmissions/\ac{E2E} delay. This is intuitively because when blockage is avoided, with high probability, large \ac{SNR} drops are not experienced. Thus, even if a signal is not correctly decoded, it needs a small boost in the \ac{SNR} which can be provided by, e.g., a single retransmission. Alternatively, one can rely  only on diversity and use Type I \ac{HARQ}  with low complexity. Finally, the blockage prediction may provide seamless handover without service interruption, because the \acp{BS} may have $n\ge 1$ slots gap for handover. 

\begin{figure}
\centering
  \includegraphics[width=0.95\columnwidth]{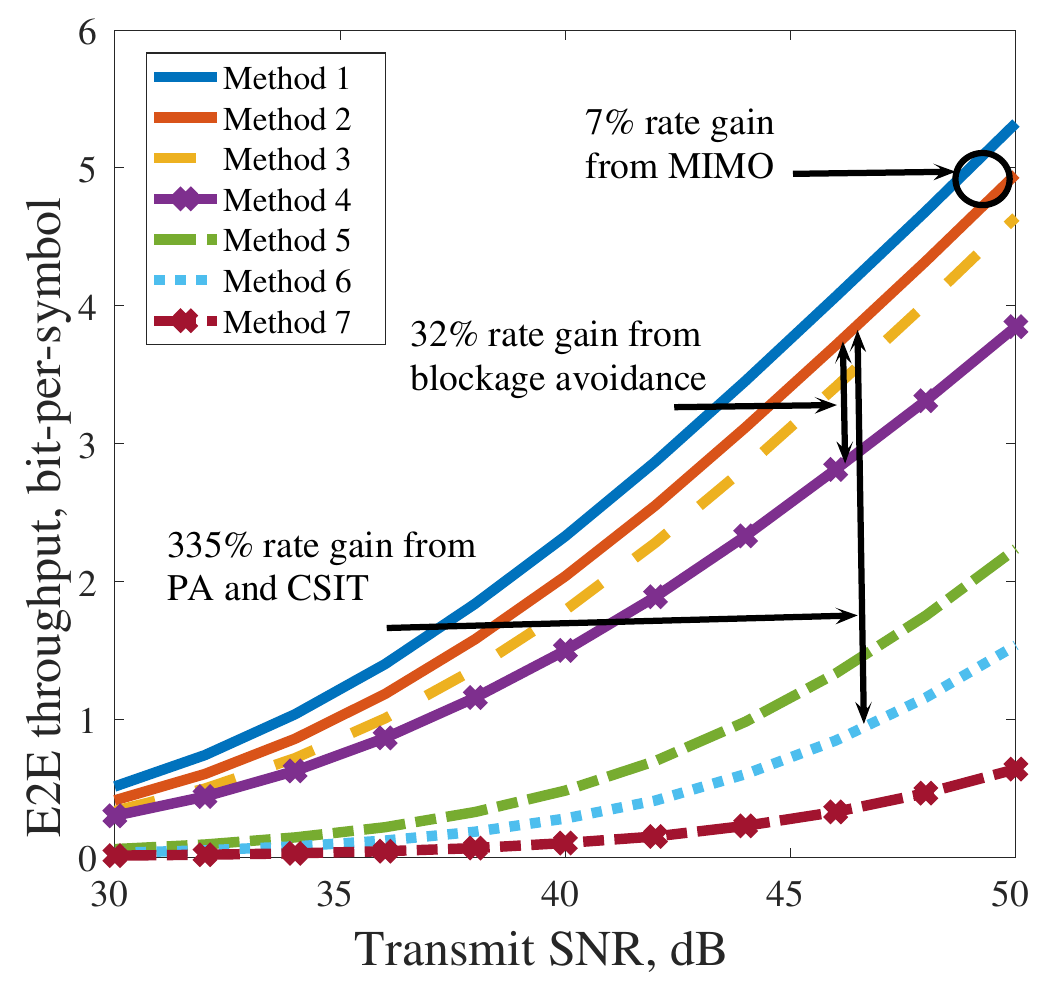}\\
\caption{\ac{E2E} throughput of the second vehicle vs transmit \ac{SNR}, 32 antennas at the \ac{BS},  carrier frequency  28 GHz,  speed 50 km/h, \ac{BS} processing delay 5 ms, codeword length 8000 symbols, antenna separation 6.6 times the wavelength and  $33\%$ blockage probability. }
\label{fig_res_1}
\end{figure}

One of the main challenges of \ac{PA} system is spatial mismatch, i.e., when the \ac{RA} does not reach the same position where the \ac{PA} (either on the same or a different vehicle) estimated the channel several time slots before. Such spatial mismatch affects the accuracy of \ac{CSIT} and, consequently, the system performance. Here, we concentrate on the highway scenario where dynamic blockage is probable. Then, \ac{LSCPA} setup is applicable as long as the vehicles are on the same lane, in a car platooning setup or with known moving speeds. However, the same approach is applicable in, e.g., trains and trams. With trams/trains, the moving trajectory is more predicable, which improves the positioning accuracy and simplifies  communication, specially at high frequencies. Moreover, beam width adaptation and the implementation of multiple antennas, as foreseen in, e.g., mobile \acp{IAB} \cite{madapatha2020integrated}, reduce the sensitivity of the \ac{PA} system to spatial mismatch at \ac{mmW} band. Finally, note that communications between
neighbouring mobile units, normally carried out via sidehaul links, is out of our scope, and we concentrate on \ac{V2I} communications. This is because we concentrate on \ac{BS} cooperation where how to acquire/share the \ac{CSIT} and handover between \acp{BS} is determined in a central unit, e.g., a \ac{BS}. However, the concept of the proposed scheme can be partly extended to V2V systems as well.



\begin{figure}
\centering
  \includegraphics[width=0.99\columnwidth]{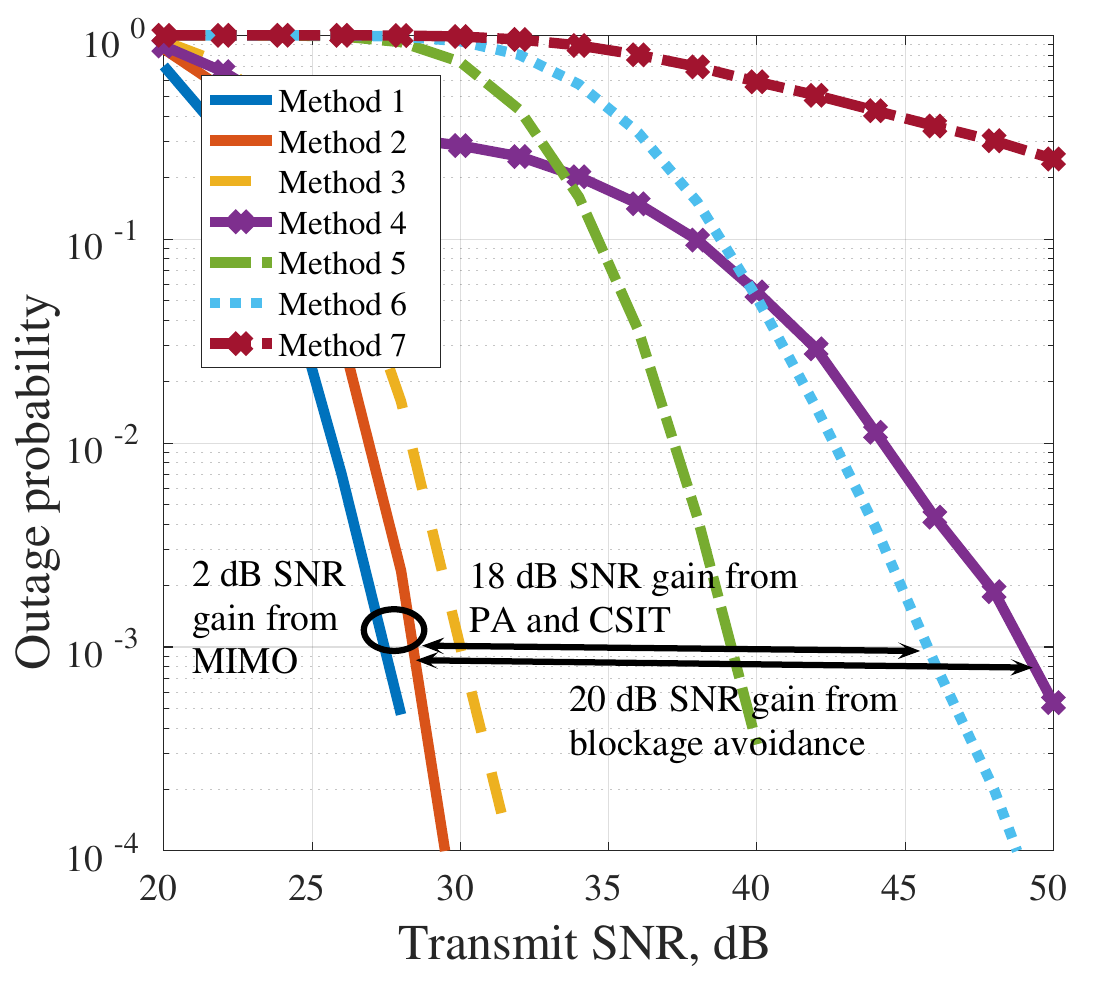}\\
\caption{Outage probability  versus transmit \ac{SNR}, 32 antennas at the \ac{BS}, $33\%$ blockage probability, carrier frequency  28 GHz, speed  50 km/h, \ac{BS} processing delay 5 ms (affecting the cases with spatial mismatch), codeword length 8000 symbols, and antenna separation 6.6 times the wavelength. The outage threshold, i.e., the minimum data rate required by the \acp{UE}, is  0.07 bit-per-symbol.}
\label{fig_res_2}
\end{figure}

Considering the setup of Fig. \ref{fig1} with two vehicles, two \acp{BS} and \ac{mmW} spectrum, Figs. \ref{fig_res_1}-\ref{fig_res_2} show the \ac{E2E} throughput  and the  outage probability of the second vehicle in the cases with 33\% blockage probability. We present the results for both  cases with and without spatial mismatch.  In the cases with no spatial mismatch, one may consider an adaptive-delay transmission scheme where the transmission delay is dynamically adapted, as a function of the speed/antennas distance, such that the \ac{BS} sends the data to the \ac{RA} at exactly the same position as the  \ac{PA} performing \ac{CSIT} acquisition. Here, \ac{CSIT} is perfect, at the cost of extra transmission delay. Note that  the adaptive-delay scheme is applicable only for a range of speeds limited by the \ac{BS}’s minimum required processing delay (see Fig. \ref{fig_res_4}). With a non-adaptive delay method, on the other hand, one can always consider the \ac{BS}’s minimum processing delay. This setup, which is more suitable for slotted communications, is at the cost of possible spatial mismatch/CSIT inaccuracy  which can be modeled as a function of the speed/spatial mismatch \cite{guo2021predictor}. Also, to evaluate the effect of \ac{MIMO} communications, considering 32 antennas at the \ac{BS}, we present the results for both cases with (one \ac{RA}, one \ac{PA}) and (two \acp{RA}, two \acp{PA}) setups at the \acp{UE}.

We compare the performance of our proposed scheme, represented in Methods 1-3, with benchmark alternative techniques of Methods 4-7:

\begin{itemize}
        \item \textit{Method 1: \ac{CP}-\ac{MIMO}-perfect \ac{CSIT}}. With \ac{MIMO} setup and \ac{MRT} \ac{BF}, two \acp{PA} and two \acp{RA}, i.e., a total of four antennas, are deployed on top of each vehicle, and, using the information provided by the front vehicle,  \ac{CP} scheme is used to avoid blockages. Here, we consider perfect \ac{CSIT} at the \acp{BS}, i.e., adaptive-delay scheme, at the cost of extra transmission delay. Finally, the blockage is avoided by switching to the closest non-blocked \ac{BS}.
        \item \textit{Method 2: \ac{CP}-MISO-perfect \ac{CSIT}}. Here, we consider the same setup as in Method 1, except that each vehicle is equipped with only two antennas; one RA and one PA, i.e., the \ac{BS}-\ac{RA} link is MISO (S: Single).
        \item \textit{Method 3: \ac{CP}-MISO-mismatched \ac{CSIT}}. Considering  \ac{CP} \acp{BS} to avoid blockage, a non-adaptive delay method is considered, based on the \ac{BS} minimum processing delay, at the cost of imperfect small-scale \ac{CSIT}.
        \item \textit{Method 4: \ac{NCP}-MISO-perfect \ac{CSIT}}. Here, we study the case with \ac{NCP} transmission from a single \ac{BS} at the cost of possible blockages. However, in each vehicle, small-scale fading is  predicted using the self \ac{PA}.
        \item \textit{Method 5: \ac{CP}-MIMO-no \ac{CSIT}}. With a total of two antennas on top of each vehicle, we deploy \ac{CP} setup to avoid blockage. However,  no small-scale CSIT is considered and  random \ac{BF} is performed while both antennas at the vehicle are used for data reception. This is a benchmark showing the gain of small-scale \ac{CSIT} acquisition.
        \item \textit{Method 6: \ac{CP}-MISO-no \ac{CSIT}}. The setup is similar to Method 5 with only one antenna at the vehicle.
        \item \textit{Method 7: \ac{NCP}-MISO-no \ac{CSIT}}. The  worst-case benchmark with no \ac{CP} framework/\ac{CSIT}, and random \ac{BF}.
\end{itemize}

Here, transmit \ac{SNR} is defined as the transmit power of the \ac{BS} normalized with unit-variance noise, and the details of the parameter settings are given in the figure captions.  \ac{E2E} throughput is defined as the average number of correctly decoded information bits per the \ac{E2E}  delay. The \ac{E2E}  delay is given by the transmission delay plus the possible processing delay at the BS. The results are presented for spatially-correlated Rayleigh fading conditions  using the Jake's correlation model.  In Figs. \ref{fig_res_1}-\ref{fig_res_2}, we study the case that, while the information provided by the front vehicle is used for blockage avoidance, the small-scale  \ac{CSIT} is obtained from the \ac{PA}(s) in self vehicle.  The potential of using the small-scale \ac{CSIT} from the front vehicle is studied in Fig. \ref{fig_res_5}.

\begin{figure}
\centering
  \includegraphics[width=1.0\columnwidth]{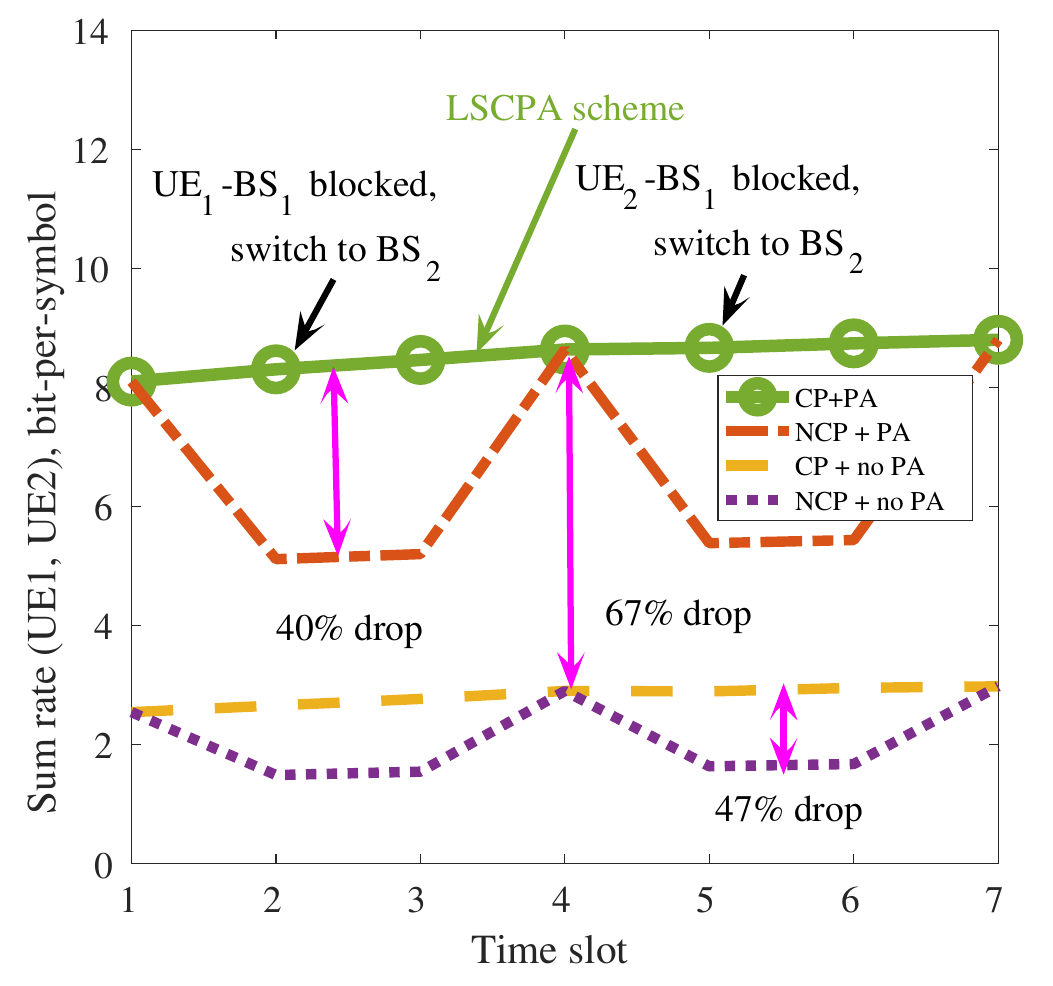}\\
\caption{An example of the sum throughput of $\text{UE}_1$ and $\text{UE}_2$ in different time slots, 32 antennas at the \acp{BS}, carrier frequency 28 GHz, transmit \ac{SNR} 50 dB, channel bandwidth 50 MHz with noise figure -174 dBm/Hz and antenna separation 6.6 times the wavelength. }
\label{fig_res_3}
\end{figure}

According to Figs. \ref{fig_res_1}-\ref{fig_res_2}, blockage predication and avoidance via \acp{BS} cooperation leads to considerable \ac{E2E} throughput and outage probability improvement, at the cost of overhead for acquisition and exchange of information about the vehicles speed/CSI. For instance, considering the cases with single \ac{PA} and \ac{RA} on each vehicle and 46 dB transmit \ac{SNR}, blockage avoidance results in $32\%$ throughput increment, compared to the cases with \ac{NCP} transmission (Fig. \ref{fig_res_1}). Also, with an outage probability  $10^{-3}$ and the parameter settings of Fig. \ref{fig_res_2}, the proposed \ac{LSCPA} scheme gives 20 dB \ac{SNR} gain, compared the \ac{NCP} scheme experiencing possible blockages, even in the presence  of perfect small-scale \ac{CSIT}. Finally, using \ac{PA} to predict small-scale \ac{CSIT} could reach a 335\% gain in \ac{E2E} throughput  at 46 dB transmit \ac{SNR} as well as 18 dB \ac{SNR} gain with an outage probability $10^{-3}$.

In summary, blockage  (resp. small-scale \ac{CSIT}) prediction is the main performance booster in terms of outage probability (resp. throughput). Hence, the proposed LSCPA method, with both blockage and \ac{CSIT} prediction can be an enabler for \textit{uninterrupted high-rate} \ac{IoV} communications. Finally, although increasing the number of antennas at the vehicles improves the achievable rate/reliability, for a broad range of parameter settings, the relative gain of \ac{MIMO} communication is marginal in the cases with perfect CSIT. With no/partial \ac{CSIT}, however, \ac{MIMO} setup leads to considerable outage probability/throughput improvement (Figs. \ref{fig_res_1}-\ref{fig_res_2}).


To demonstrate the \ac{LSCPA} procedure in detail, considering an $\text{UE}_0-\text{UE}_1-\text{UE}_2$ car platooning setup, Fig. \ref{fig_res_3} presents the sum throughput of $\text{UE}_1$ and $\text{UE}_2$ following  $\text{UE}_0$, possibly used as a predictor, within seven sample time slots.  We present the result for four cases, namely, 1) \ac{LSCPA} using self \ac{PA} for \ac{CSIT}  acquisition, 2) non-\ac{LSCPA} using self \ac{PA} for \ac{CSIT} acquisition, 3)  \ac{LSCPA} but no \ac{PA} \ac{CSIT} and random \ac{BF}, and 4) no blockage avoidance/\ac{CSIT}. As seen in Fig. \ref{fig_res_3}, the blockage deteriorates the system performance,  during $T_2-T_3$ and $T_5-T_6$  where $\text{UE}_1$ and $\text{UE}_2$ are blocked by a truck, respectively, resulting in almost $40\%$  throughput drop with  small-scale \ac{CSIT} prediction. For both cases with and without blockage detection, the lack of small-scale \ac{CSIT} leads to significant throughput reduction, for instance, $67\%$ throughput drop with blockage avoidance. However, deploying \ac{LSCPA} the system experiences an \textit{almost constant} \ac{QoS} and the  throughput is improved, compared to the considered benchmark schemes (the slight throughput increase in these sample slots is because of the UEs getting closer to the BSs).

\begin{figure}
\centering
  \includegraphics[width=1.0\columnwidth]{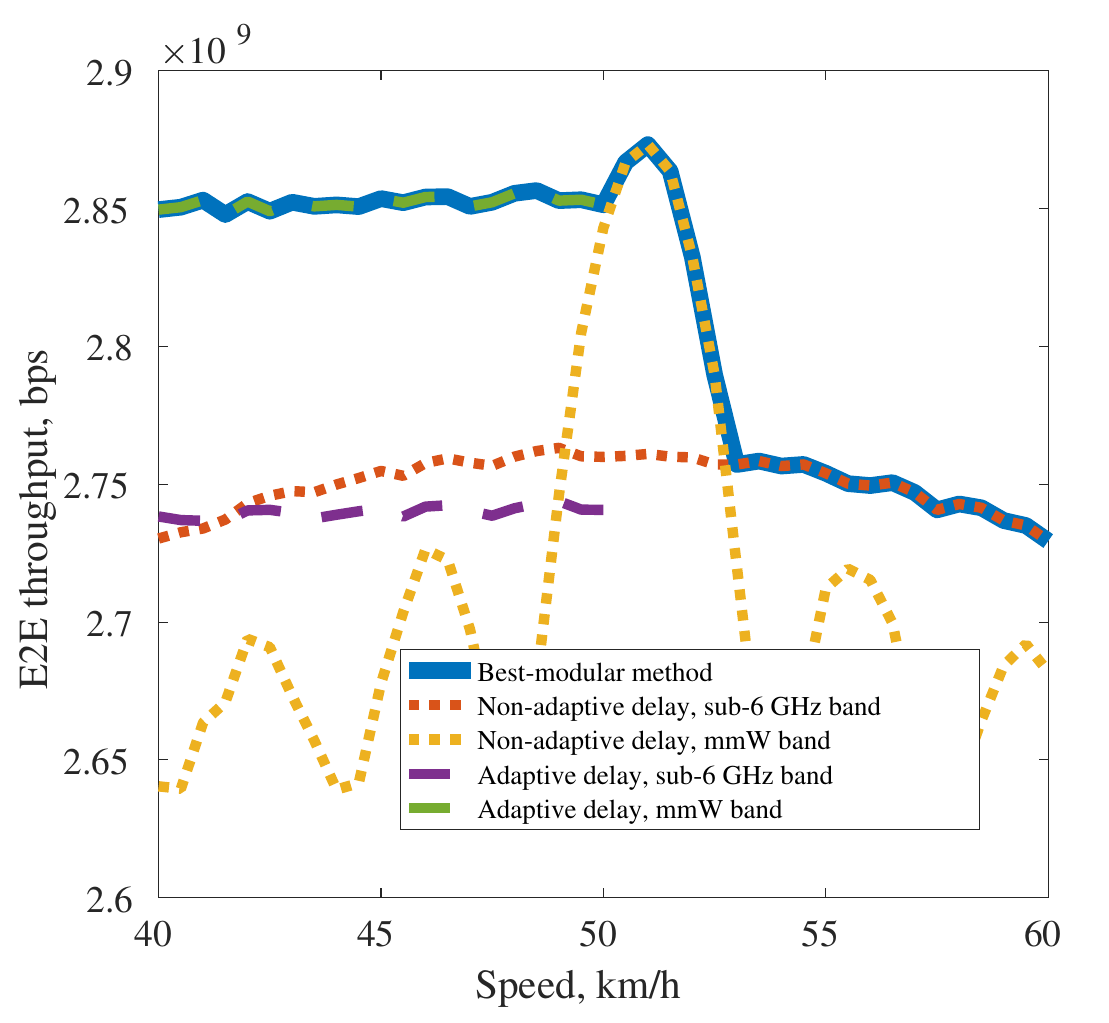}\\
\caption{\ac{E2E} throughput in bit-per-second (bps) as a function of the speed. We consider both 2.8 and 28 GHz frequencies and 32 antennas at the \ac{BS}. We set the transmit power to 50 dBm, noise spectrum density as -174 dbm/Hz, and channel bandwidth for \ac{mmW} (resp. sub-6 GHz) as 80 (resp. 70) MHz, codeword length 10000 symbols, antenna separation for \ac{mmW} and sub-6 GHz are 6.6 and 0.6 times the wavelength, respectively.  }
\label{fig_res_4}
\end{figure}

The performance of the \ac{LSCPA} method depends on the considered carrier frequency and delay adaptation capability. For this reason, Fig. \ref{fig_res_4} studies the effect of spatial mismatch, considering both 2.8 and 28 GHz without blockage. Here, the \ac{E2E} throughput is demonstrated for both  adaptive- and non-adaptive-delay methods.

Compared to 2.8 GHz, the throughput is more sensitive to spatial mismatch and speed variation at 28 GHz (Fig. \ref{fig_res_4}). On the other hand, adaptive-delay method shows robust performance in various carrier frequencies. However, the feasibility of adaptive-delay scheme is limited within a speed range, e.g., less than 50 km/h in Fig. \ref{fig_res_4}, determined by the \ac{BS} minimum required processing delay. Importantly, there is not a single method providing the maximum throughput; at low speeds, utilizing adaptive-delay method at \ac{mmW} spectrum leads to maximum throughput. At moderate speeds, however, the highest throughput is achieved by exploiting  spatial correlation and using  non-adaptive-delay scheme at 28 GHz. Finally, at high speeds, where the sensitivity to spatial/\ac{BF} mismatch increases, the maximum throughput is achieved via non-adaptive-delay method operating at 2.8 GHz.

Note that, although adaptive-delay method at gives the highest throughput for a range of speeds, from a network perspective, it may not be of interest as it may introduce unplanned interference to the network. Moreover, in practice cellular networks have a limited transmission time
interval granularity which may limit the efficiency of the adaptive-delay scheme.

\begin{figure}
\centering
  \includegraphics[width=1.0\columnwidth]{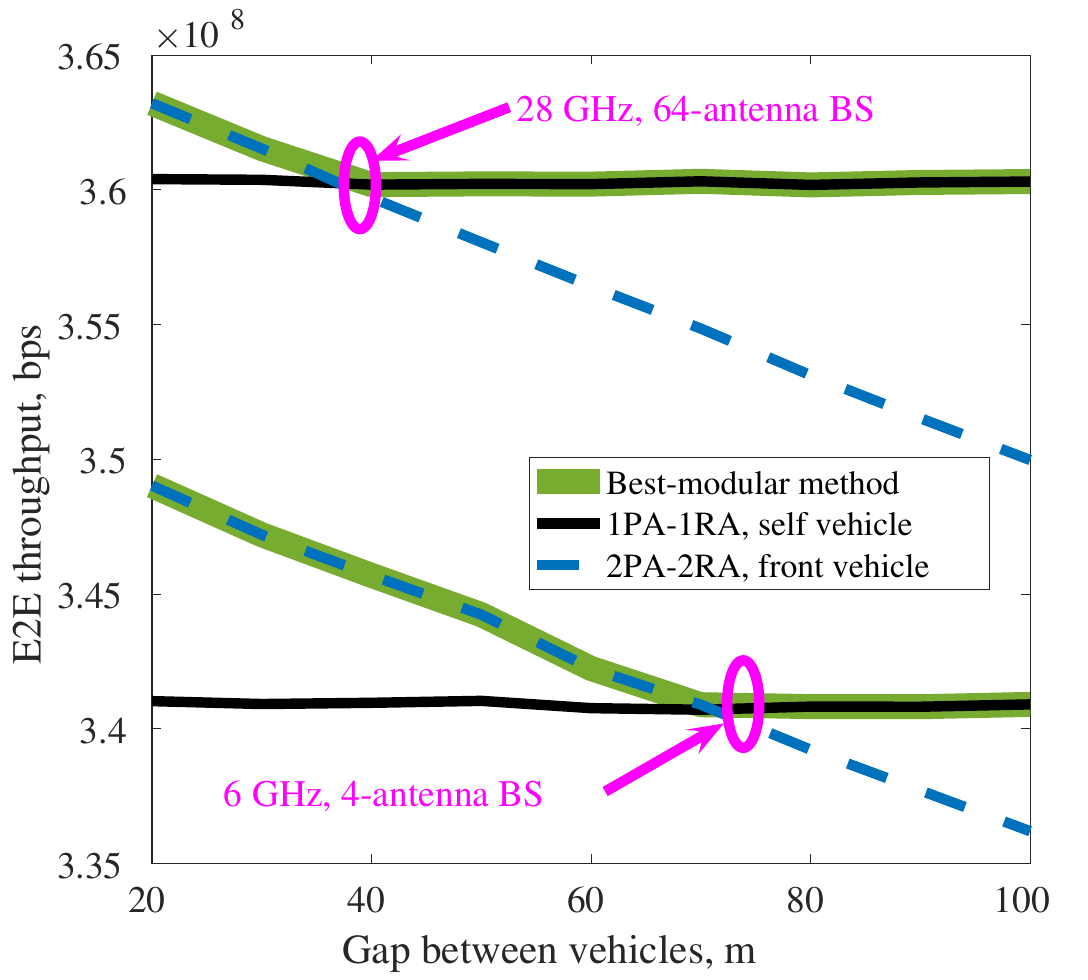}\\
\caption{\ac{E2E} throughput  as a function of the gap between the front and behind vehicles. Carrier frequencies are set to  6 and 28 GHz. Numbers of antennas at the \ac{BS} are 64 for \ac{mmW} and 4 for sub-6 GHz. The channel bandwidth is set to 10 MHz for both cases, vehicle speed 80 km/h, delay at the \ac{BS}  5 ms, transmit power 40 dBm, and the codeword length  8000 symbols.}
\label{fig_res_5}
\end{figure}
Whether to use the self or the front vehicle \ac{PA} for small-scale prediction depends on, e.g., the carrier frequency, speed and number of antennas. Figure \ref{fig_res_5} elaborates on this point where, considering both 28 and 6 GHz bands and two antennas per vehicle, we study the  throughput versus the gap between the vehicles for the following alternative schemes:

\begin{itemize}
    \item One PA, one RA, self-vehicle \ac{PA}: As in typical PA scheme \cite{guo2021predictor}, the front antenna is used only for channel estimation required for data transmission to the behind \ac{RA}.
    \item Two \acp{RA}, front vehicle \ac{PA}: Here, the \ac{BS} obtains the \ac{CSIT} from the front vehicle. Thus, using adaptive-delay scheme, both antennas of behind vehicle are utilized for data reception while the transmission parameters are adapted based on \ac{CSIT} achieved through front vehicle with a delay penalty.
\end{itemize}

As seen, at \ac{mmW} band the highest throughput is achieved by utilizing the self-vehicle \ac{PA}, unless for the cases with small gap between the vehicles where exploiting the front vehicle \ac{CSIT} and using both antennas of the behind vehicle for data reception improves the throughput. At 6 GHz, however, exploiting the \ac{CSIT} provided by the front vehicle is useful for a broad range of gaps between the vehicles. In summary, there is a trade-off between utilizing both antennas for data reception and the extra delay for data transmission due to utilizing the front vehicle \ac{CSIT}, and the maximum throughput is achieved in a modular setup where the front vehicle (resp. the self-vehicle \ac{PA}) \ac{CSIT} is used in the cases with small (resp. large) gap between the vehicles.   In practice, handling the spatial mismatch by utilizing the \ac{CSIT} from the front vehicle requires good alignment along the moving direction. This can be achieved with cm-level by, for example, coordinated control over platooning setups, advanced location methods \cite{6g2020white}, or different coaches in trains/trams, whereas \ac{mmW}-level localization are involved in ongoing research projects such as EU Hexa-X.

\section{Standardization Progress for IoV}

Since Rel. 16, 3GPP is deeply involved in standardizations enabling \ac{IoV}. To complement LTE V2X, 3GPP Rel. 16 developed a new V2X standard based on 5G \ac{NR}  air interface. Here, different levels of automation levels ranging from no automation (level 0) to full automation (level 5) and various car platooning, advanced driving, extended sensors and remote driving use cases are defined (see \cite{garcia2021CST} for details). Particularly, with advanced driving and extended sensor use-cases, the vehicles share the data obtained from their own sensors with the surrounding vehicles/infrastructure, to improve the perception of the environment beyond that obtained by vehicles' own sensors. Also, to better serve the mobile nodes, dual connection-based mobility management and \ac{UE}-based handover are defined. These can be enablers for practical
implementation of the \ac{LSCPA} method.
 
From relaying perspective, 3GPP Rel. 16-17 have specified standardizations for \ac{IAB} as the main relaying method in 5G \ac{NR} \cite{madapatha2020integrated}. Here, although mobile \ac{IAB} were not considered, a large part of the specifications are compatible with mobility. More importantly, Rel. 18 \ac{IAB}, to start in early 2022, will be fully dedicated to mobile \ac{IAB}/vehicle-mounted relays. In that case, different issues such as movement of mobile \ac{IAB} between different central units, dynamic adaptation of routing tables, interference management, etc., need to be handled.  With both V2X and \ac{MR} functionalities, the sensitivity of the \ac{mmW} narrow \ac{BF} to inaccurate \ac{CSIT}/\ac{BF} mismatch and blockage should be carefully taken care of, where, along with other methods, the \ac{LSCPA} concept may be an attractive candidate.

\section{Conclusions}
Aiming for uninterrupted IoV communications, we demonstrated the potentials of utilizing the front vehicles information for dynamic blockage and small-scale fading prediction in high-speed links. Introducing the ongoing standardization attempts enabling IoV communications, we verified the effect of MIMO transmission and different carrier frequencies on the network performance, where the best E2E throughput is obtained in a modular setup of different carrier frequencies/data transmission techniques, depending on the vehicles distances/speeds. Our simulations  show that the proposed LSCPA concept is a potential solution to support future high-speed IoV links. However, there is still room for theoretical/testbed evaluations identifying the potentials and challenges of LSCPA.

\section*{Acknowledgement}
This work was supported in part by VINNOVA (Swedish Government Agency for Innovation Systems) within the VINN Excellence Center ChaseOn, and in part by the European Commission through the H2020 Project Hexa-X under Grant 101015956.

\bibliographystyle{IEEEtran}
\bibliography{main.bib}

\begin{IEEEbiography}{Hao Guo} [S'17] (hao.guo@chalmers.se)  is a postdoc in Chalmers, Sweden.
\end{IEEEbiography}

\begin{IEEEbiography}{Behrooz Makki} [M'19, SM'19] works as Senior Researcher in Ericsson Research, Sweden. 
\end{IEEEbiography}

\begin{IEEEbiography}{Mohamed-Slim Alouini} 
[S'94, M'98, SM'03, F'09] is a Professor in King Abdullah University of Science and Technology, Saudi Arabia. 
\end{IEEEbiography}

\begin{IEEEbiography}{Tommy Svensson} [S’98, M’03, SM’10] is a Full Professor in Chalmers, Sweden.
\end{IEEEbiography}
\end{document}

%% file: main_acro.tex
\acrodef{AoA}{angle-of-arrival}
\acrodef{AoD}{angle-of-departure}
\acrodef{BS}{base station}
\acrodef{BF}{beamforming}
\acrodef{BPL}{building penetration loss}
\acrodef{bps}{bit-per-second}
\acrodef{RA}{receive antenna}
\acrodef{CP}{cooperative}
\acrodef{NCP}{non-cooperative}
\acrodef{PA}{predictor antenna}
\acrodef{CDF}{cumulative distribution function}
\acrodef{DFT}{discrete Fourier transform}
\acrodef{DU}{distributed unit}
\acrodef{DSRC}{dedicated short-range communication}
\acrodef{HARQ}{hybrid automatic repeat request}
\acrodef{PDF}{probability density function}
\acrodef{CU}{centralized unit}
\acrodef{c.u.}{channel use}
\acrodef{SNR}{signal-to-noise ratio}
\acrodef{SINR}{signal-to-interference-plus-noise ratio}
\acrodef{CSIT}{channel state information at the transmitter}
\acrodef{CSI}{channel state information}
\acrodef{bps}{bits per second}
\acrodef{npcu}{nats per channel use}
\acrodef{E2E}{end-to-end}
\acrodef{MRC}{maximum ratio combining}
\acrodef{MT}{mobile terminal}
\acrodef{MRT}{maximum ratio transmission}
\acrodef{OFDM}{orthogonal frequency division multiplexing}
\acrodef{LoS}{line-of-sight}
\acrodef{LiDAR}{light-detection-and-ranging}
\acrodef{LSCPA}{large-scale cooperative PA}
\acrodef{NLoS}{non-line-of-sight}
\acrodef{NLoSb}{non-line-of-sight, building}
\acrodef{NLoSbm}{non-line-of-sight, building, muted antenna}
\acrodef{NLoSv}{non-line-of-sight, vehicle}
\acrodef{MIMO}{multiple-input multiple-output}
\acrodef{MISO}{multiple-input single-output}
\acrodef{MRN}{moving relay node}
\acrodef{MR}{moving relay}
\acrodef{mmW}{millimeter wave}
\acrodef{UE}{user equipment}
\acrodef{4G}{fourth generation}
\acrodef{5G}{fifth generation}
\acrodef{LTE}{Long-Term Evolution}
\acrodef{LTE-A}{Long-Term Evolution-Advanced}
\acrodef{UL}{uplink}
\acrodef{DL}{downlink}
\acrodef{QoS}{quality-of-service}
 \acrodef{IoV}{Internet-of-Vehicle}
\acrodef{TDD}{time division duplex}
\acrodef{FDD}{frequency division duplex}
\acrodef{GPS}{Global Positioning System}
\acrodef{IAB}{integrated access and backhaul}
\acrodef{SIMO}{single-input-multiple-output}
\acrodef{SISO}{single-input-single-output}
\acrodef{ZF}{zero-forcing}
\acrodef{VPL}{vehicle penetration loss}
\acrodef{V2I}{vehicle-to-infrastructure/network}
\acrodef{V2X}{vehicle-to-everything}
\acrodef{V2V}{vehicle-to-vehicle}
\acrodef{3GPP}{The 3rd Generation Partnership Project}
\acrodef{URLLC}{ultra-reliable low-latency communication}
\acrodef{NR}{new radio}
\acrodef{DAPS}{dual active protocol stack}
\acrodef{OVT}{on-vehicle transceiver}